\documentclass[11pt]{article}

\usepackage{booktabs}
\usepackage{tabularx}
\usepackage{amsmath}
\usepackage{url}
\usepackage[breaklinks]{hyperref}
\usepackage{breakurl}

\newcommand{\eat}[1]{}

\usepackage{graphicx}

\usepackage{multirow}
\usepackage[authoryear]{natbib}
\usepackage{rotating}
\usepackage{bbm}
\usepackage{latexsym}

\newcommand{\captionfonts}{\normalsize}

\makeatletter  
\long\def\@makecaption#1#2{%
  \vskip\abovecaptionskip
  \sbox\@tempboxa{{\captionfonts #1: #2}}%
  \ifdim \wd\@tempboxa >\hsize
    {\captionfonts #1: #2\par}
  \else
    \hbox to\hsize{\hfil\box\@tempboxa\hfil}%
  \fi
  \vskip\belowcaptionskip}
\makeatother   
\begin{document}

\title{Floating-Point Multiplication\\ Using Neuromorphic Computing}
\author{Karn Dubey \hspace{0.5in} Urja Kothari \hspace{0.5in} Shrisha Rao}
\date{}
\maketitle

\begin{abstract}
Neuromorphic computing describes the use of VLSI systems to mimic
neuro-biological architectures and is also looked at as a promising
alternative to the traditional von Neumann architecture.  Any new
computing architecture would need a system that can perform
floating-point arithmetic. In this paper, we describe a neuromorphic
system that performs IEEE 754-compliant floating-point multiplication.
The complex process of multiplication is divided into smaller
sub-tasks performed by components Exponent Adder, Bias Subtractor,
Mantissa Multiplier and Sign OF/UF.  We study the effect of the number
of neurons per bit on accuracy and bit error rate, and estimate
the optimal number of neurons needed for each component.
\end{abstract}

\noindent{\bf Keywords:} IEEE 754, floating point arithmetic,
neuromorphic computing, Neural Engineering Framework (NEF)

\section{Introduction}

Neuromorphic computing has recently become prominent as a possible
future alternative to the traditional Von Neumann
architecture~\citep{vonneuman} of computing.  Some of the problems that
are commonly faced when working with classical CMOS-based Von Neumann
machines are the limitations on their energy efficiencies, and also
the absolute limits to speed and scaling on account of physical
limits~\citep{cmead,koch1999}.  Though Moore's Law held for long and
made possible rapid and sustained progress in hardware
performance~\citep{moore}, it is now quite clear that this will not
last. Hence, there is a need to look for alternative computing
architectures, including neuromorphic
computing~\citep{wang2017,kim2015,esser2016}. The Von Neumann
architecture also has an inherent problem, commonly called the ``Von
Neumann bottleneck,'' because of the limited bandwidth between the CPU
and the main device memory.  Thus, newer architectures often avoid a
wide gap between processing and main memory~\citep{dmonroe,moore}.

Rapid growth in cognitive applications is one of the important
motivations for interest in neuromorphic computing, which promises the
ability to perform a high number of complex functions through parallel
operation.  Neural solutions are possible for machine learning
problems that involve complex mathematical
calculations~\citep{eliasmith2013,brain}.  There have been some
attempts to develop systems of computation on neuromorphic
architectures~\citep{koch1999,goasmann2016} but not much has been done
in the specific area of numerical computations, particularly for
floating-point arithmetic.

Floating-point arithmetic~\citep{ieee-floating754} is ubiquitous in
scientific as well as general computing.  It is a basic operation that
should be supported by any computational architecture.  In this paper,
we describe a system which can perform the multiplication of two IEEE
754-compliant floating-point numbers on a neuromorphic architecture.
Our work is an extension to~\citet{IEEE754} who showed how floating
point addition can be achieved using neuromorphic computing.  We have
designed a modular architecture which performs the conventional
multiplication process~\citep{Mark2019}, but instead of logic gates it
uses groups of neurons as the basic unit.  The architecture is easily
scalable to double-precision floating point numbers.

The system is designed on the basis of the Neural Engineering
Framework (NEF) which, as the name suggests, provides a basic
framework to develop a neuromorphic system.  For the implementation,
simulation and testing of our design we used
Nengo~\citep{nengo,bekolay2013}, a graphical and scripting-based
software package for simulating large-scale neural systems.  To use
Nengo, we define groups of neurons called \emph{ensembles}, and then
form connections between them based on what
computation~\citep{AddExampleNengo,MulExampleNengo} should be
performed.

The architecture is divided into four components: Exponent Adder, Bias
Subtractor, Mantissa Multiplier, and Sign/Overflow and Underflow.  The
Exponent Adder uses a stage-wise adder which takes 8-bit exponents and
produces an 8-bit output along with carry.  The Bias Subtractor takes
the output of the Exponent Adder and subtracts the bias and produces
8-bit output.  The subtraction is done using 2's complement method.
The Mantissa Multiplier is the core of our system design; it follows a
stage-wise process, taking two 23-bit mantissa inputs, and outputs a
23-bit resultant mantissa (see Section~\ref{mantissa_multiplication}).
Our system also indicates if there is an overflow or underflow during
the exponent addition process (see Section~\ref{S/OF}).

We used two performance analysis metrics: Mean Absolute Error (MAE)
and Mean Encoded Error (MEE) to estimate the performance of our
system.  We have also observed the effect on accuracy by varying
number of neurons of each component in our system.

The rest of the paper is structured as follows. We first give a brief
description of the IEEE 754 floating-point multiplication process in
Section~\ref{Section_Floating_Point}, and then briefly describe the
Neural Engineering Framework (NEF) and its three basic principles:
representation, transformation and dynamics, in
Section~\ref{Section_NEF}.  After this we explain the overall
architecture in Section~\ref{system_architecture} using
Figure~\ref{architecture_IEEE_floating_point}.  The performance
analysis metrics in Section~\ref{Section_Observation_Result} deal with
the two metrics that we have used to evaluate our system: the Mean
Absolute Error (MAE) and Mean Encoded Error (MEE).  In
Section~\ref{Accuracy_v/s_number_of_neurons} we describe the
relationship between the number of neurons and accuracy, and in
Section~\ref{Section_Bit_Error} we describe the relationship between
the number of neurons and bit error. In
Section~\ref{Section_Total_Numer_Neurons} we describe how we estimated
the optimal number of neurons required for all the ensembles, and list
them in Table~\ref{table1}. Finally, we present the conclusions of our
work in Section~\ref{Section_Conclusion}.

\section{Background}\label{background}

First we briefly discuss the floating-point multiplication process as
per the IEEE 754 standard~\citep{Mark2019}, then we describe the Neural
Engineering Framework (NEF) which we have used to design, simulate and
evaluate our system~\citep{tstewart}.

%%%%%%%%%%%%%%%
\begin{figure}%[htbp]
\centering{\includegraphics[width=0.8\textwidth]{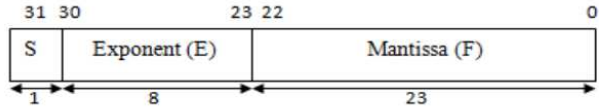}}
\caption{IEEE 754 32-bit floating-point representation}
\label{floating_point_representation}
\end{figure}

\subsection{IEEE 754 floating-point multiplication}\label{Section_Floating_Point}

\begin{figure}%[htbp]
\centering
\includegraphics[width=0.4\textwidth]{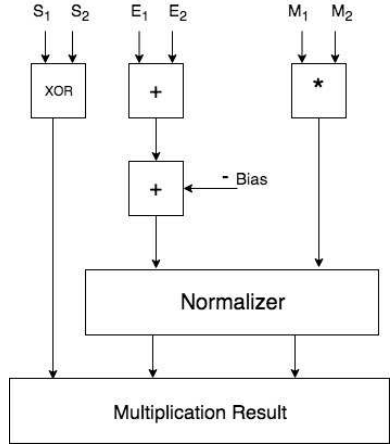}
\caption{Process for multiplication of floating point numbers}
\label{floating_point_multiplication}
\end{figure}

Figure~\ref{floating_point_multiplication} illustrates the overall
process of multiplication of two floating-point numbers
\textit{Input\textsubscript{1}} and \textit{Input\textsubscript{2}}
represented in binary format.
Figure~\ref{floating_point_representation} is an example of how a
32-bit floating-point number is represented according to the IEEE 754
standard~\citep{ieee-floating754}. A sign bit is used to represent
whether the number is positive or negative. 8 and 23 bits are used to
represent the exponent and mantissa values respectively.  While
designing this system we assumed that both inputs, i.e., the two
floating-point numbers, are represented according to the IEEE 754
standard in binary representation.

In Figure~\ref{floating_point_multiplication}, The exponents
E\textsubscript{1} and E\textsubscript{2} are added. The Bias value
(127) is subtracted from the sum of E\textsubscript{1} and
E\textsubscript{2}. The difference is placed in the Exponent field
(see Figure~\ref{floating_point_representation}). Each mantissa is of
24 bits (23 bits + 1 hidden bit). Mantissa M\textsubscript{1} and
M\textsubscript{2} are multiplied and give a 48 bit output; if the
48\textsuperscript{th} bit is 1 then the result is normalized by right
shifting and incrementing the resultant exponent (if it is 0, then
nothing further is to be done).  To find the resultant mantissa, we
take the first 24 bits (23 bits + 1 hidden bit). The resultant sign
field is the XOR of the two sign bits S\textsubscript{1} and
S\textsubscript{2}.

For a better understanding of the above algorithm, see~\citet{Yi2009}.

\subsection{Neural Engineering Framework}
\label{Section_NEF}

The Neural Engineering Framework
(NEF)~\citep{tstewart,voelker2017a,voelker2017b} is a computational
framework which is used for mapping computations to the biological
network of spiking neurons.  It provides a general way to generate
circuits that have analytically determined synaptic weights to provide
the desired functionality. NEF consists of three principles:
representation, transformation, and dynamics~\citep{nengo,C.Eliasmith}.
Using these principles we can implement NEF for constructing complex
neural models.

\subsubsection{Representation} \label{'Representation'}

Neural representations are defined by the combination of nonlinear
encoding and weighted linear decoding.  (We use the notation given
by~\citet{tstewart}.)  If $x$ is the value represented by a neural
ensemble and $e_i$ is the encoding vector for which that neuron fires
most strongly, then activity $a_i$ for each neuron can be represented
as follows:
\begin{equation}
    a_{i}=G_{i}[\alpha_{i}\mathbf{e}_{i}\cdot \mathbf{x}+b_{i}],\ \ i=1\ldots n 
\end{equation}
    
    where $G$ is neural non-linearity, $\alpha_i$ is the gain
    parameter, and $b_i$ is the constant background bias current for
    the neuron.  Given an activity, estimating the value of $x$ can be
    done by finding a linear decoder $d_{i}$.
    \begin{equation}
         \hat{x}=\sum{a_{i}d_{i}}
    \end{equation}
   
    Decoding weights $d_{i}$ can be seen as a least-squares
    minimization problem, as $d_{i}$ is set of weights that minimizes
    the difference between $x$ and its estimate~\citep{tstewart}.
    \begin{equation}
         d=\Gamma^{-1}\Upsilon
    \end{equation}
   \begin{equation}
       \Gamma\textsubscript{ij}=\Sigma_{x}a_{i}a_{j}
   \end{equation}
    \begin{equation}
        {\Upsilon_j}^{}=\Sigma_{x}a_{j}x
    \end{equation}

\subsubsection{Transformation}

Section~\ref{'Representation'} shows how to encode and decode a vector
in the distributed activity of a population of neurons.  To perform
computation, these neurons need to be connected and information needs
to be transferred from one group of neurons to another.  This is done
via synaptic connections.  In other words, we want our connections to
compute some functions.  Transformation is used for approximation of
these functions~\citep{tstewart}.  Transformation is another weighted
linear decoding for approximating function $f(x)$; the decoded weights
$d^{f(x)}$ can be computed as:
     \begin{equation}
         d^{f(x)}=\Gamma^{-1}\gamma^{f(x)}
     \end{equation}
     \begin{equation}
         \Gamma{ij}=\Sigma_{x}a_{i}a_{j}
     \end{equation}
    \begin{equation}
         {\Upsilon_j}^{f(x)}=\Sigma_{x}a_{j}f(x)
    \end{equation}
    
In general, the more non-linear and discontinuous function is, the
lower is the accuracy of its computation.  Accuracy also depends on
other factors like neuron properties, number of neurons, and the
encoding method. The NEF is using the same trick seen in support
vector machines~\citep{N.Christian} to allow complex functions to be
computed in a single set of connections as we choose $e_{i}$,
$\alpha_{i}$ and $b_{i}$. The function $f(x)$ is constructed by a
linear sum of tuning curves of neurons, so a wider variety of tuning
curves leads to better function approximation~\citep{tstewart}.

\subsubsection{Dynamics}

Dynamics of the neural systems can also be modeled in NEF using
control-theoretic state variables.  However, NEF also provides a
direct method for computing dynamic functions of the form:
\begin{equation}
    \frac{dx}{dt} = F(x) + H(u)
\end{equation}
 where $x$ is the value getting represented, $u$ is some input, and
 $F$ and $G$ are some arbitrary functions.

\section{System Architecture}\label{system_architecture}

\begin{figure}%[htbp]
\centering
  \includegraphics[width=\textwidth]{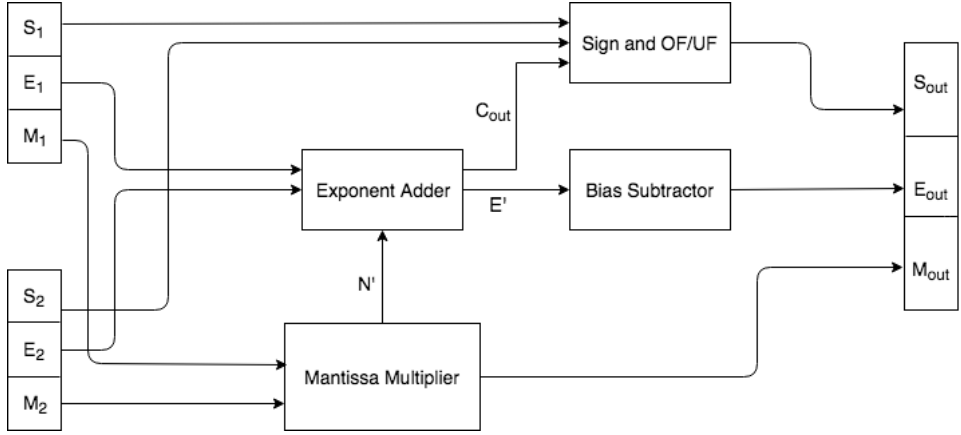}
\caption{Architecture diagram for single precision IEEE floating point number multiplication}
\label{architecture_IEEE_floating_point}
\end{figure}

We have designed a system that performs floating-point multiplication
according to the IEEE
standard~\citep{ieee-floating754}. Figure~\ref{architecture_IEEE_floating_point}
illustrates the system architecture. The two inputs are represented as
(${S}_1$,${M}_1$,${E}_1$) and (${S}_2$,${M}_2$,${E}_2$) and the output
is represented as
($S_{\mathrm{out}}$,$M_{\mathrm{out}}$,$E_{out}$). Here $S_{i}$
represents the sign bit, $M_{i}$ represents the mantissa bit, and
$E_{i}$ represents the exponent bit, where $i \in \{1,2, \ldots,
\mathrm{out}\}$. This representation follows the IEEE-754 32-bit
floating point standard~\citep{ieee-floating754}.  Each of the
components is described in the following subsections.

\subsection{Simulation}\label{Simulation}

For simulation we use the Leaky Integrate-and-Fire (LIF) neural
model. We create the neural ensembles using the Nengo library to
represent input information.  The values of two properties,
\emph{radius} and \emph{dimension} of the ensemble are set in the same
way as~\citet{IEEE754}.  We have also used the same
encoding scheme as~\citet{IEEE754} to transfer the output
of one ensemble as an input to another ensemble.  For the AND ensemble
(Section~\ref{mantissa_multiplication}) we have used the following
encoding scheme:
\begin{equation}\label{encoding}
        E(\hat{x}_i)= 
\begin{cases}
    1, & \hat{x}_i \geq 1.5\\
    0, & \text{otherwise}
\end{cases}
\end{equation}

\subsection{Exponent Adder} \label{Exponent_Adder}

As shown in Figure~\ref{architecture_IEEE_floating_point}, the
Exponent Adder takes three inputs: $E_1$, $E_2$ and a normalization
bit produced by the Mantissa Multiplier (see
Section~\ref{mantissa_multiplication}). It performs addition of 8-bit
$E_1$,$E_2$ and Normalization bit (as $C_{in}$) produces an 8-bit
output $E'$ and a carry bit $C_{\mathrm{out}}$.  To implement this
stage-wise addition process, we construct a network that takes two
inputs (the corresponding bits of two exponents, i.e., $a_i$ and
$b_i$, where $0 \leq i \leq 7$, and represent them using two different
ensembles, say A\_ensemble and B\_ensemble. These two ensembles are
then connected to another ensemble, say C\_ensemble, through synaptic
connections. Now the sum of A\_ensemble and B\_ensemble is represented
by C\_ensemble.  The adder is implemented in same way as in prior
literature~\citep{IEEE754,AddExampleNengo}.  The $C_{\mathrm{out}}$ bit
produced by the Exponent Adder is used in the calculation of overflow
and underflow (see Section~\ref{S/OF}).
\begin{figure} %[htbp]
\centering
\includegraphics[width=0.6\textwidth]{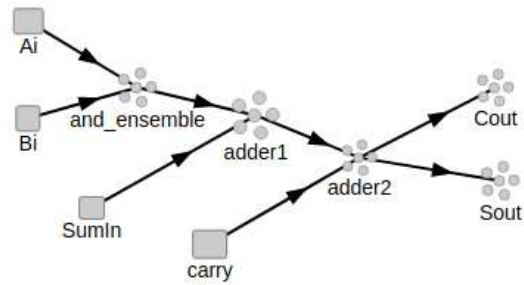}
\caption{Building block of Mantissa Multiplier component consisting of AND Ensemble and Adder}
\label{build_block_dig}
\end{figure}
\subsection{Mantissa Multiplier} \label{mantissa_multiplication}

\begin{figure*} \label{mul_block_dig}
\centering
\includegraphics[width=0.8\textwidth]{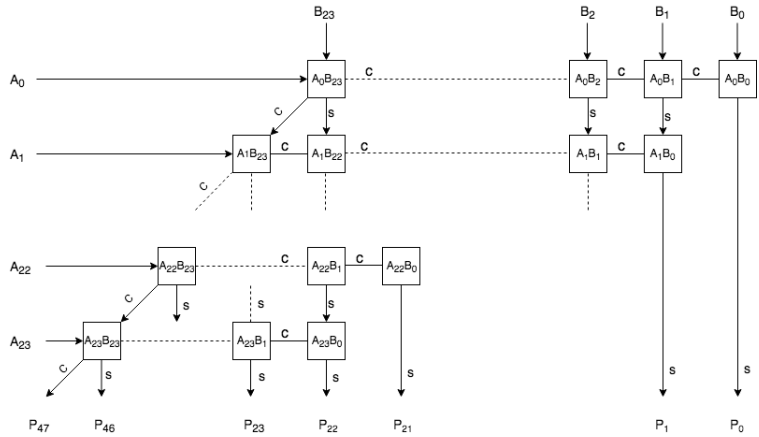}
\caption{Process for multiplication of floating point numbers}
\label{process_mul}
\end{figure*}
The Mantissa Multiplier component is the core of our system.  It is a
stage-wise process.  Figure~\ref{process_mul} shows its working.  We
use an AND ensemble and adders as building blocks for multiplication
(see Figure~\ref{build_block_dig}).  The AND Ensemble is used to
implement neuromorphic AND logic. The encoding scheme for it is given
in~(\ref{encoding}).  In the AND ensemble we connect two inputs.  If
both inputs are 1 then the output is more than 1.5, so the output is
set to 1; otherwise it is 0.  The working and connection of each block
at every stage is described below in detail by taking two mantissa $A$
and $B$:

\begin{itemize}
    \item Each block $j$ of stage $i$ is given four inputs $A_i$,
      $B_j$, sum $s_\mathrm{in}$ produced by block $(j+1)$ of stage
      ${(i-1)}$ and carry $c_\mathrm{in}$ from block $(j-1)$ of stage
      $i$, where $ 0 \leq i,j \leq 23$.
    \item As shown in Figure~\ref{process_mul}, the last block of each
      stage $i$ takes $c_{\mathrm{out}}$ of the previous stage's last
      block as $s_{\mathrm{in}}$.
    \item The AND Ensemble of each block of every stage performs AND
      operation on $A_{i}$ and $B_j$ and outputs $A_iB_j$.
    %\item Takes in two bits \textit{a} and \textit{b} as input,
    %performs AND operation and produces \textit{a}\textit{b}.
    \item The adder of blocks performs 3-bit addition of $A_{i}B_{j}$,
      $s_\mathrm{in}$ and $c_\mathrm{in}$ and produces
      $s_{\mathrm{out}}$ and
      $c_{\mathrm{out}}$~\citep{IEEE754,AddExampleNengo}
    \item $s_{\mathrm{out}}$ and $c_{\mathrm{out}}$ produced as
      outputs are fed as input to the next stage and next block
      respectively.
    
\end{itemize}

The first block of every stage is given $c_{\mathrm{in}}$ as $0$.  The
output obtained at each stage ensemble is encoded and fed to the next
stage ensemble as input. Encoding of the output at each stage helps to
filter and boost up the output signal.  At each stage the first
block's $s_{\mathrm{out}}$ represents the output bit of the mantissa
as shown in Figure~\ref{process_mul}. At the end of this process we
get a 48-bit product.  If the 48\textsuperscript{th} bit is 1, then we
set the normalization bit, right shift the product by one, which
thereby results in incrementing the exponent by one (see
Section~\ref{Exponent_Adder}).  The resultant product is in the 1.M
form as per IEEE standard. We take the first 23 bits from M and stores
it as a resultant mantissa $M_{\mathrm{out}}$.

\subsection{Bias Subtractor} \label{bias_subtraction}

As shown in Figure~\ref{architecture_IEEE_floating_point}, this
component subtracts the bias from the result which we get from
exponent addition.  The subtraction is done using the 2's complement
method~\citep{davidj}. This is achieved by taking the 2's complement
of the bias and then performing addition.  To perform 2's complement,
we design a converter, which takes 8-bit bias and represents it
using a neural ensemble.  We take a 1's complement of bias by flipping
its bits, and then take the 8-bit adder and add 1 to 1's
complement of bias.  The final output is stored as a resultant
exponent $E{\mathrm{out}}$.

\subsection{S\textsubscript{out} and OF/UF} \label{S/OF}

This component computes $S_{out}$ bit of the output along with OF/UF
(overflow/underflow) flag which can then be used for rounding.  It
computes output sign bit $S_{\mathrm{out}}$ by performing a
neuromorphic XOR operation on two sign bits $S_{1}$ and
$S_{2}$~\citep{IEEE754}. Overflow is indicated by setting the OF/UF
flag as 1 if a carry is found during exponent addition.
\begin{figure}%[htbp]
\centering
\includegraphics[width=0.6\textwidth]{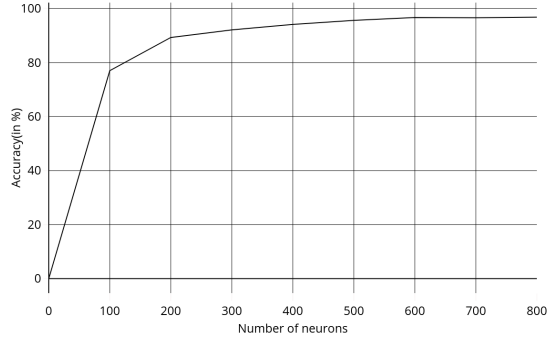}
\caption{Accuracy vs. Number of neurons/Ensemble Graph of Mantissa Multiplier}
\label{accuracy_mantissa_mult}
\end{figure}

\section{Observations and Results}
\label{Section_Observation_Result}

We simulated the individual components of the system and integrated
them to arrive at fully functional IEEE floating point multiplication.
We probed the outputs of each component at a time interval of 10ms and
computed errors in each of them.  We used the following two techniques
for evaluating the performance of each component.

\[\mathrm{Mean \ Absolute \ Error} = \frac{{\sum|\mathrm{Computed \ val} - \mathrm{Actual \ val}|}}{\mathrm{number \enspace of \enspace values}}\]

\[\mathrm{Accuracy} = (1 - \mathrm{Mean \ Absolute \ Error}) \times 100\]

The Mean Absolute Error is the measure of the absolute difference
between the actual bit value and the value computed by our system,
averaged over all the bits.  In our case MAE obtains due to
approximating a discontinuous function using NEF, plus noise and
randomness in spiking neurons.

\[\mathrm{Mean \ Encoded \ Error} =\displaystyle\frac{\sum|\mathrm{Actual \enspace bit} \mathbin{\oplus} \mathrm{Encoded\enspace val}|}{\mathrm{number\enspace of \enspace bits}}\]

We encoded the output value of each component and compare it with
actual bit value. In other words we calculated hamming distance
between the encoded bit value and actual bit value then averaged it
over all the bits.

\subsection{Accuracy versus number of neurons} \label{Accuracy_v/s_number_of_neurons}

Figure~\ref{accuracy_mantissa_mult} %and~\ref{accuracy_exponent_adder}
illustrates the accuracy of the Mantissa Multiplier. (For the Bias
Subtractor and Exponent Adder we get very similar graphs.)

We varied the number of neurons starting from 100 to maximum of 800
per bit, and observed the accuracy across all components.  We observed
that the accuracy initially increases with the number of neurons but
after some threshold value of neurons, increase in accuracy is not
significant.  In the Mantissa Multiplier component we can see that
accuracy increases rapidly until the number of neurons reach 300;
after that there is no significant improvement.

\eat{

\begin{figure}%[htbp]
\includegraphics[width=9cm, height=5cm]{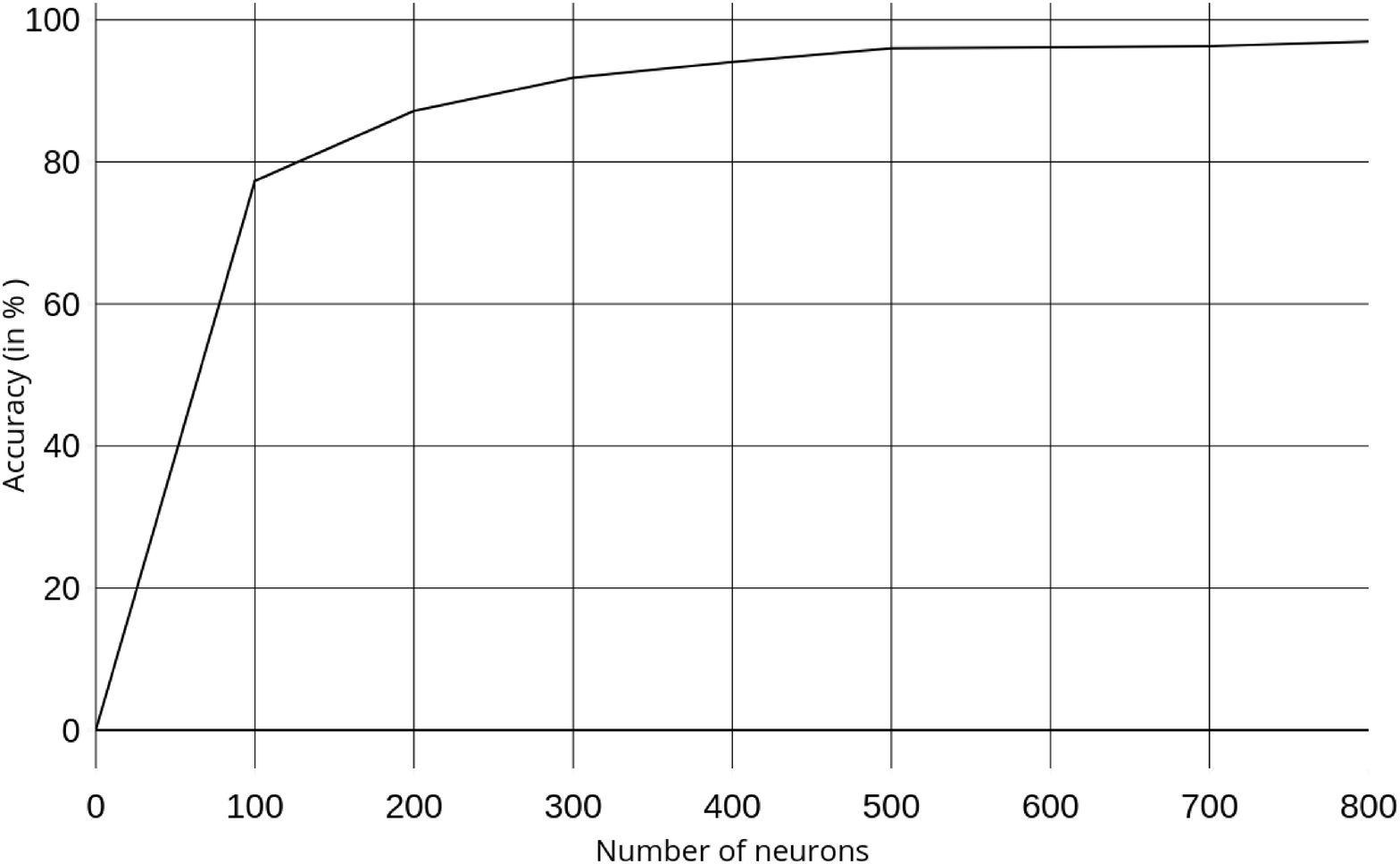}
\caption{Accuracy vs Number of Neurons/Ensemble Graph of Bias Subtractor}
\label{accuracy_exponent_adder}
\end{figure}

}

\subsection{Bit error v/s number of neurons:}
\label{Section_Bit_Error}

For each Mantissa Multiplier component we observed that bit error is
high when the number of neurons is very low.  In the Mantissa
Multiplier, when the number of neurons are below 200, we got 1 bit
error out of 48 bits which is roughly equivalent to 2\%.  After
increasing the number of neurons to 300 we get no bit errors.  For the
Exponent Adder and Bias Subtractor we get no bit errors even for
number of neurons below 200.

\subsection{Total number of neurons}
\label{Section_Total_Numer_Neurons}

We observed in Section~\ref{Accuracy_v/s_number_of_neurons} that the
accuracy increases with an increase in the number of neurons. We
estimated the optimal number of neurons required in all for all
ensembles, as in Table~\ref{table1}

%\eat{

\begin{table} 
\caption{Number of neurons for each ensemble} \label{table1}
\begin{center}
\begin{tabular}{l c} 
\toprule
Component & Number of neurons \\ 
\midrule
Exponent Adder & 300 \\ 
\midrule
Bias Subtractor & 300 \\ 
\midrule
Mantissa Multiplier & 600 \\
\midrule
Sign and OF/UF & 100\\
\bottomrule
\end{tabular}
\end{center}
\end{table}

%}

\section{Conclusion}
\label{Section_Conclusion}

In this paper we describe an approach to build an IEEE-754 standard
floating point unit using neuromorphic hardware with spiking neurons.
Such devices can mimic aspects of the brain's structure, and may be an
energy-efficient alternative to the classical Von Neumann
architecture.  Such a neuromorphic floating-point unit is a critical
step in developing an alternative, neuromorphic CPU architecture.

Our architecture comprises a complex floating-point multiplication
process. The most complex part of the process is the Mantissa
Multiplier, which we have realized successfully by using stage-wise
multiplication and a robust encoding scheme. The architecture is
easily scalable to double-precision floating point numbers also.  We
have checked the presence of overflow and underflow errors which than
can be handled separately.  We have studied the affect of number of
neurons on accuracy and bit error.  Finally we derive the optimal
number of neurons required for each component, giving an indication of
the hardware resources required to implement this approach.

\end{document}